\documentstyle[11pt,aaspp4]{article}
\def\lap{\hbox{${_{\displaystyle<}\atop^{\displaystyle\sim}}$}}
\def\gap{\hbox{${_{\displaystyle>}\atop^{\displaystyle\sim}}$}}

\begin{document}

\title{Constraints on the Production of Ultra-High-Energy Cosmic Rays
by Isolated Neutron Stars}

\author{ Aparna Venkatesan,  M. Coleman Miller,\footnote{Compton GRO
Fellow}  and Angela V. Olinto } \affil{Department of Astronomy and
Astrophysics, and Enrico Fermi Institute,} \affil{ University of
Chicago, 5640 South Ellis Avenue, Chicago, IL 60637}

\begin{abstract}

The energetics, spectrum, and composition of cosmic rays with
energies below $\sim 10^{15}$ eV are fairly well explained by
models involving supernova shocks.  In contrast, no widely
accepted theory exists for the origin of ultra-high energy
cosmic rays (UHECRs), which have energies above $10^{15}$
eV. Instead of proposing a specific model, here we place strong
constraints on any model of UHECRs involving isolated neutron
stars (no companions). We consider the total power requirements
and show that the only viable power source associated with
isolated neutron stars is rotation. Mechanisms based on
accretion from the interstellar medium fall short of the
necessary power despite the most optimistic assumptions.  Power
considerations also demonstrate that not enough rotational
energy is tapped by a ``propeller"-like acceleration of
interstellar matter.  The most promising source of energy is
rotational spindown via magnetic braking.  We examine
microphysical energy loss processes near magnetized neutron
stars and conclude that the most likely site for yielding UHECRs
from isolated neutron stars is near or beyond the light cylinder.

\end{abstract} 

\section{Introduction}

The energy spectrum of cosmic rays is well established between
$\sim 10^8$ eV and $\sim 10^{20}$ eV (Axford 1994; Bird et
al. 1994, and references therein). There is clearly a ``knee" at
about $10^{15}$ eV, at which the spectrum changes from
$N(E)\,\sim E^{-2.7}$ to $N(E)\,\sim E^{-3.1}$, and an ``ankle''
centered on $10^{18.5}$ eV, beyond which $N(E)\,\sim
E^{-2.7}$. Cosmic rays of energy up to the knee are widely
accepted as originating in shocks associated with galactic
supernova remnants, but supernova shocks have difficulties
producing particles of higher energy. It is therefore necessary
to posit another process to produce these ultra-high energy
cosmic rays (UHECRs) beyond the knee of the spectrum. Cosmic
rays with energies above $\sim 10^{19}$ eV are generally thought
to be extragalactic (Axford 1994; Bird et al. 1994), although
they may also originate in an extended halo of the Galaxy
(Vietri 1996).

Recent attention has focused on isolated neutron stars (without
any companions) as promising sites for high-energy phenomena,
such as X-ray and UV radiation (Blaes \& Madau 1993; Madau \&
Blaes 1994, hereafter BM93 and MB94, respectively), as well as
ultra-high-energy gamma rays (see, e.g., Harding 1990).  In
particular, it has been suggested that accretion from the
interstellar medium by isolated neutron stars may provide the
necessary energetics and spectrum up to cosmic ray energies of
$\sim$ $10^{15}$ eV (Shemi 1995).  Here we analyze in detail the
prospects for these isolated neutron stars to be the source of
cosmic rays above $10^{15}$ eV.  We narrow down the set of
allowed models by requiring first that any model be able to
generate the total power observed in UHECRs, then by examining
microphysical energy loss processes to determine the maximum
energy to which cosmic rays may be accelerated around neutron
stars.

The plan of this paper is as follows. In \S\ 2 we review the
data on the energy and spectrum of cosmic rays above the knee.
The energy generation rate of cosmic rays depends on both the
observed spectrum and the dependence of confinement time on
energy.  The latter is uncertain past $\sim 10^{12}$ eV
(M\"uller et al. 1991), but we show that at least $10^{38}$ erg
s$^{-1}$ of cosmic rays beyond $10^{15}$ eV leave the Galaxy.
This is thus the energy generation rate of any viable mechanism.
In \S\ 3 we consider specific power sources associated with
isolated neutron stars, including magnetic fields, kinetic
energy, accretion from the interstellar medium, and rotation.
We find that only rotation produces the required power per
neutron star; in particular, accretion from the interstellar
medium is too weak.  We show further that a ``propeller"
mechanism (Illarionov \& Sunyaev 1975) is not likely to convert
rotational energy to UHECRs, because ram pressure from the
neutron star wind prevents accretion until the rotational energy
is too low to account for UHECRs.  If rotational energy is
transformed into UHECRs, it is likely to do so directly from the
neutron star wind.  In \S\ 4 we consider microphysical loss
processes near magnetized neutron stars to determine the maximum
energy to which a particle may be accelerated.  We find that
synchrotron and curvature radiation are the most significant
loss processes, but that if acceleration takes place near the
light cylinder the losses may be negligible.  We also show that
the preferential escape of higher energy particles may occur
beyond the light cylinder. Finally, in \S\ 5 we discuss our
results and summarize the viable mechanisms for the production
of UHECRs by isolated neutron stars.

\section{Power and Spectral Requirements}

The energy generation rate of cosmic rays in the Galaxy above a
given energy $E_0$ is $$P( \geq E_0) \propto \int_{E_0}^{\infty}
N(E)EE^{\eta} dE \ . $$ Here $N(E)$ is the differential number
distribution of observed cosmic rays, and the factor $E^\eta$
accounts for the dependence of the galactic confinement time on
the cosmic rays' energy. Empirically, $N(E) \propto
E^{-\gamma}$, with a spectral index $\gamma=2.7$ for $E\lap
10^{15}$ eV and $\gamma=3.1$ for $10^{15}$ eV $\lap E \lap
10^{18.5}$ eV.  There is evidence that the spectrum hardens
again to $\gamma$ $\sim$ 2.7 above $\sim 10^{18.5}$ eV (Bird et
al. 1994, and references therein).

Cosmic rays of higher energy have larger gyroradii, and thus
escape more easily from the Galaxy, than cosmic rays of lower
energy. Above $\sim 10^{19}$ eV cosmic rays are no longer
confined by the galactic magnetic field, so $\eta\rightarrow 0$;
below this energy the original source spectrum of cosmic rays
will be flattened. To understand this effect, consider that the
power necessary to explain the observed luminosity in cosmic
rays with energy above $E_0$ is $P(\geq E_0) \propto \tau_{\rm
conf}^{-1}$ (Milgrom \& Usov 1996), where $\tau_{\rm conf}$ is
the confinement time. We assume here that $\tau_{\rm conf}$
depends on energy as $E^{-\eta}$.  On theoretical grounds,
Biermann (1993) estimated $\eta \sim 0.3$ for the energy range
$10^{13}$ eV to 3 $\times$ $10^{18}$ eV. However, from
measurements of the relative abundance of secondaries in the
cosmic ray spectrum and the predicted escape rates at a given
energy, M\"uller et al. (1991) determined that $\eta\simeq 0.6$
between $10^{10}$ eV and $10^{12}$ eV. Outside this energy
range, there are no empirical determinations of $\eta$.

\begin{figure}[htb]
\plotone{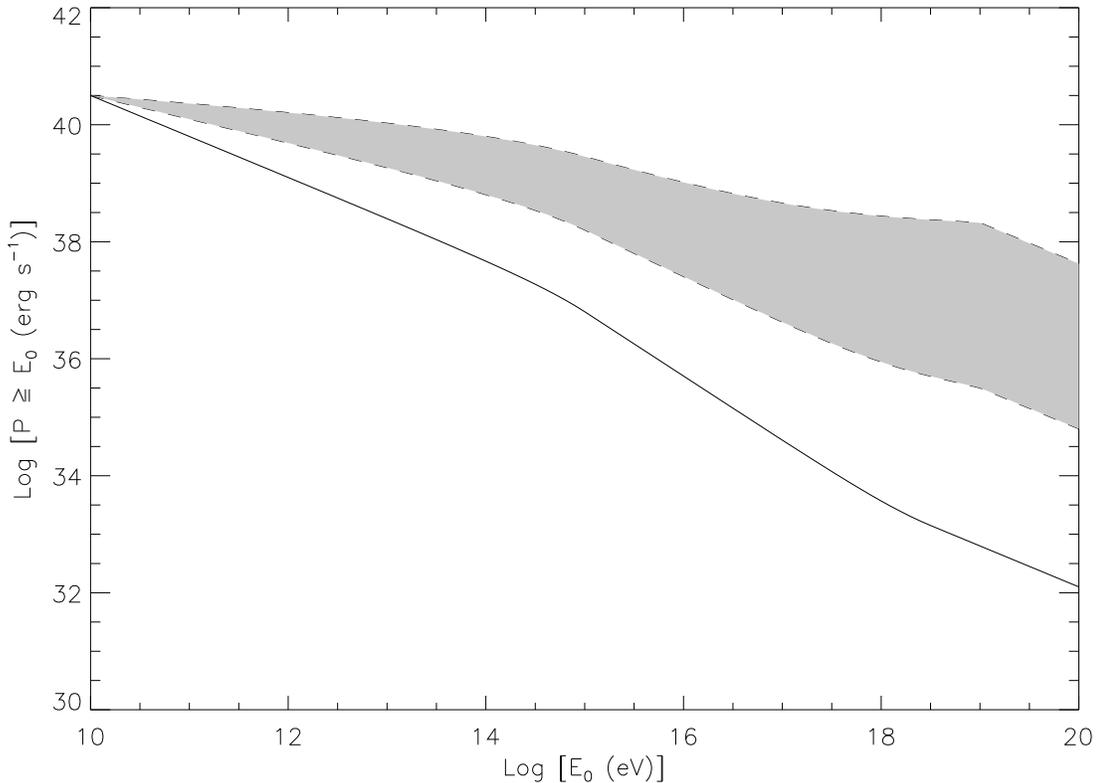}
 
\caption{Power requirements as a function of the energy of
cosmic rays. Shaded area is bounded from below by $\eta = 0.3$
and from above by $\eta = 0.6$.  Solid line is for $\eta = 0$.}

\end{figure}

In Figure 1, we plot the power requirement as a function of
energy from $10^{10}$ eV to $10^{20}$ eV, taking the uncertainty
in $\eta$ into account. The shaded region of the graph
represents the range in the power for possible values of $\eta$,
with the lower bound drawn for $\eta=0.3$ and the upper bound
for $\eta=0.6$.  The solid curve has $\eta$ = 0; confinement
effects are also neglected for $E_0$ $>$ $10^{19}$ eV. Each
curve is normalized by setting $P(\geq 10^{10}\,{\rm
eV})=10^{40.5}$ erg s$^{-1}$ (Milgrom \& Usov 1996).  From this
curve we find that the minimum energy generation rate in the
Galaxy past $10^{15}$ eV is $\sim 10^{38}$ erg s$^{-1}$, and so
this is the minimum power that must be met by models seeking to
explain UHECRs.

\section{Power Sources}

Since neutron stars are produced in supernovae at a maximum rate of
one per ten years, or one every $10^{8.5}$ seconds, they have to
produce at least $10^{38}\times 10^{8.5}=10^{46.5}$ ergs per neutron
star in UHECRs during their lifetimes in order to account for cosmic
rays above the knee. 

Neutron stars have many sources of energy, but some of them are
not promising as sources of cosmic rays.  For example, if their
average magnetic field is $\lap 10^{13}$ G, as is inferred for
all rotationally powered pulsars, then each neutron star has a
total magnetic energy $\lap 10^{43}$ ergs, many orders of
magnitude too low.  Observations of pulsar birth velocities (see
below) tell us that the kinetic energy of neutron stars could
easily account for cosmic rays; at a median space velocity of
500 km s$^{-1}$ the energy per neutron star is $\approx 4\times
10^{48}$ ergs.  To tap that energy the star must be slowed down
by collisions.  However, collisions between stars are extremely
improbable in the disk of the Galaxy, and typical interstellar
medium (ISM) densities are not sufficient to slow down neutron
stars over the age of the universe.

We are therefore left with two main energy sources, which we
consider below: accretion and rotation.  In this paper we
consider only isolated neutron stars, so by accretion we mean
accretion from the ISM. Although accretion has been proposed as
a promising source of cosmic rays, we find that the power that
can be generated is insufficient to account for UHECRs, let
alone all cosmic rays of energy above $10^{10}$ eV. Therefore,
rotation is the only viable power source.

\subsection{Accretion from the ISM}

Recent work has suggested that neutron stars accreting from the
ISM may be sources of UV and X-ray radiation (BM93;
MB94). Typically, Bondi-Hoyle accretion (Bondi \& Hoyle 1944) is
considered, which becomes much less efficient with increasing
neutron star velocities.  Our understanding of the velocity
distribution of neutron stars has been revised significantly in
the last few years; prior to 1993, their average velocity was
assumed to be $\sim 150$ km s$^{-1}$.  However, Lyne \& Lorimer
(1994) showed that after selection effects were removed, this
number increased to $\sim 450$ km s$^{-1}$.  Frail, Goss, \&
Whiteoak (1994) inferred that neutron stars associated with
supernova remnants have average velocities of $\sim 500$ km
s$^{-1}$.  These results greatly lessen the overall power from
ISM-accreting neutron stars, as compared to earlier estimates.

Bondi-Hoyle accretion is the most efficient accretion possible
onto the star from the ISM and, therefore, gives an upper limit
to the luminosity. A neutron star in the disk of the Galaxy with
a spatial velocity of $200 \, v_{200}$ km s$^{-1}$, moving
through an ISM of average number density $ n_0$ cm$^{-3}$ has a
Bondi-Hoyle accretion rate of ${\dot M}_{BH}=3.6 \times 10^8 \;
n_0 \; v_{200}^{-3}$ g s$^{-1}$.  The energy released by this
accretion cannot exceed the free-fall energy on the stellar
surface, which gives $\sim 2\times 10^{20}$ erg g$^{-1}$, or a
luminosity of $L_{ff}\approx 6.8\times 10^{28}\; n_0 \;
v_{200}^{-3}$ erg s$^{-1}$.  An average neutron star, accreting
for the entire lifetime of our Galaxy ($3\times 10^{17}$
seconds), will therefore produce only $\sim 2 \times 10^{46}$
ergs via accretion from the ISM, taking $n_0 \sim v_{200} \sim
1$.  Despite the most optimistic assumptions, this is still too
small to explain the observed power in UHECRs.

A realistic estimate of the accretion power will yield much less
than $ 2 \times 10^{46}$ ergs, if for example, centrifugal
barriers and ram pressure of the neutron star wind are taken
into account (BM93, and references therein).  Moreover, the
efficiency of converting accretion energy into cosmic rays is
certainly much less than unity, and the preheating of the ISM by
radiation from the neutron star can also decrease the accretion
rate (Blaes, Warren \& Madau 1995). The total mass accreted by
all isolated neutron stars in the Galaxy depends strongly on the
number of low-velocity neutron stars; since the accretion rate
goes as the inverse cube of the neutron star's velocity, for
which we have assumed a lower value than the median one, our
upper limit is robust.

An individual neutron star moving slowly in a very dense environment,
e.g., in the cores of giant molecular clouds, could perhaps, produce
cosmic rays at a high rate (Shemi 1995). However, since the
accretion rate goes linearly with the ISM density, the total
accretion power onto neutron stars is proportional to the {\it
average} ISM density.  Thus, neutron stars in dense clouds cannot be
considered as typical and the overall energy constraint for UHECRs
will not be met by accretion models for isolated neutron stars.

\subsection{Rotation of Neutron Stars}

The most rapidly rotating neutron star known is PSR 1937+214,
which has an angular velocity of $\sim 4\times 10^3$ s$^{-1}$
(Becker \& Helfand 1983). Since neutron stars have moments of
inertia $I\sim 10^{45}$ erg s$^2$, a star such as PSR 1937+21
has a rotational energy $E={1\over 2}I\omega^2\sim 10^{52}$
ergs, many orders of magnitude greater than the $\sim 10^{46.5}$
ergs required per neutron star.  Thus, even if most neutron
stars rotate more slowly and the efficiency of cosmic ray
generation is much less than unity, rotation is by far the most
promising cosmic ray power source related to isolated neutron
stars.

This rotational energy may be converted to the kinetic energy of
particles, either from the interstellar medium or from the
neutron star itself.  In the first case, although the particles
are accreted from the ISM and may be accelerated by, e.g., a
propeller mechanism, we are not using power from accretion
itself, so the constraints of the previous section do not
apply. The accreted ISM may gain energy from the rotating
magnetosphere through a single encounter, producing very high
energy particles, or through shock acceleration involving many
scatterings, leading to a power-law spectrum. A potential site
for this is the light cylinder radius, which is the maximum
cylindrical radius out to which corotation with the neutron star
may be causally enforced. It is given by $r_{L} = cP/2\pi = 4.8
\times 10^9 P$ cm, where $P$ is the period of the neutron star
in seconds.  Within $r_L$, we take the neutron star's magnetic
field to have a dipole geometry ($B \sim$ $r^{-3}$), while
beyond $r_{L}$ the magnetic field is azimuthal and has a
$r^{-1}$ dependence.

However, accretion from the ISM is possible only if it is not
prevented by the pressure of the neutron star wind.  A rough criterion for
this can be derived by equating the wind pressure, ${\dot E}/4\pi r^2
c$, with the ram pressure of the infalling material, $\rho_{ISM}v^2$,
at a radius $r$ equal to the Bondi-Hoyle radius $r_{BH}=2GM/v^2$. 
Here $\rho_{ISM}$ is the density of the ISM and $v$ is the velocity
of the neutron star. The rate of rotational spin-down energy is  
$$|{\rm \dot E}| = {B^2 R^6 \omega^4 \sin^2 \alpha
\over{6c^3}}\, ,$$ where $B$ is the surface magnetic field of
the star, $R$ is the radius of the star, $\omega$ is the star's
angular velocity, and $\alpha$ is the angle between the star's
magnetic and rotational axes (see, e.g., Shapiro \& Teukolsky
1983). This emitted dipole radiation is manifested as an outward
wind of charged particles.  We find that the ram pressure
exceeds the neutron star wind pressure, and accretion occurs,
only if the neutron star has spun down to a period larger than
about $\sim$ 14.5 seconds.  This is consistent with the estimate
of BM93 that $P_B\gap 20B_{12}^{1/2}n_0^{-1/4}v_{200}^{1/2}$
seconds in order for accretion to proceed, where $B = 10^{12}
B_{12}$ G, and with the estimates of other authors (see, e.g.,
Shemi 1995 or Harding 1990).

We apply this condition to rotation-powered neutron star models
for UHECRs, where at least $10^{46.5}$ ergs are needed per neutron
star. Setting this number equal to $E_{rot} = \frac{1}{2} I
\omega^2 \sim 10^{45} \omega^2$, we find that: \begin{center}
$P_{\rm max} \simeq 1.12 \ {\rm s} \ {\rm or}$ $\omega_{\rm
min} \simeq 5.6 \ {\rm s}^{-1} \ .$ \end{center}

This criterion must be met if we want to generate cosmic rays of
energy above $10^{15}$ eV using the rotational energy of neutron
stars in the Galaxy. This is incompatible with the earlier
result for the minimum period necessary to allow accretion onto
the neutron star.  Thus, by the time that accretion can
overwhelm the neutron star wind, the neutron star's rotational
energy will be insufficient to account for the observed power in
UHECRs, if $B \gap 10^9$ G.

These difficulties can be avoided in models where particles
originate from within the star's light cylinder.  The conclusion
of this section is therefore that, on energetic considerations
alone, the only way for isolated neutron stars to be the primary
source of power for UHECRs is if that power source is the star's
rotation. Moreover, the accelerated particles cannot come from
the ISM.

\section{Composition, Acceleration, and Loss Processes}

We have shown that in viable neutron star models for UHECRs, the
particles originate from within the light cylinder, and rotation
is the source of power. We next examine whether these models can
give the observed composition of cosmic rays and accelerate
particles to the requisite energies, and if microphysical energy
loss processes produce strong cutoffs in the particles'
spectrum. After addressing some issues related to composition,
we consider specifically an acceleration mechanism that has
often been suggested as promising and involves using the
potential associated with the extremely high surface fields of
neutron stars.  We then examine energy losses, particularly
those from synchrotron and curvature emission, and, using these
results, we conclude by pointing out potentially viable sites
for cosmic ray acceleration.

\subsection{Composition}

At energies of about $10^{15}$ eV, the spectrum of cosmic rays
becomes steeper, and their composition is believed to shift from
being primarily protons below this energy to mostly heavier nuclei,
such as iron, above it (Bird et al. 1994; Gaisser et al. 1993).
There is also some evidence for a proton component to the spectrum
emerging above $10^{19}$ eV (Bird et al. 1994, and references
therein).  A model for the origin of cosmic rays with energies in
excess of $10^{15}$ eV must, therefore, at least allow the composition
to be biased towards heavy nuclei up to $10^{19}$ eV.

The ions in a neutron star wind are most likely to come from the
surface of the star.  This composition is difficult to predict,
but it will probably be biased in the direction of iron.
Simulations of fallback from Type II supernovae (which can
produce neutron stars) show that the division between matter
that escapes to infinity and matter that falls onto the central
star occurs roughly in the silicon layer (see, e.g., Timmes,
Woosley \& Weaver 1996).  If the matter accretes with a fair
fraction of its free-fall velocity, it should spall into lighter
elements, probably hydrogen and helium (Bildsten, Salpeter \&
Wasserman 1992).  However, for much of the fallback the
temperature on the surface is at least several hundred million
degrees, implying that fusion to iron will be rapid.  If
accretion from the ISM is suppressed because of the neutron star
wind, the propeller mechanism, or other reasons, then the
composition of the atmosphere should depend on the evolution of
the material accreted by fallback.  Since models of cooling
neutron stars indicate that the surface temperature remains
above $\sim 10^7$ K for a few years, the light elements may be
fused.  Thus, although the composition of the neutron star
atmosphere is by no means certain, it is plausible that it
consists mainly of iron.

\subsection{Acceleration by $E\parallel B$}

A frequently suggested mechanism for the production of high-energy
ions near neutron stars is the acceleration of those ions through
potential drops associated with strong electric fields parallel to
the neutron star magnetic field, either near the star or farther out,
near an ``outer gap" region (Cheng, Ho \& Ruderman 1986).  In
principle, the largest potential drop associated with a neutron star
with magnetic field $B$, angular velocity $\Omega$, and radius $R$ is
$$\Phi_{\rm max}= 300 {\Omega R\over c}BR \; \; {\rm V;}$$ so, for
$\Omega=10^4$ s$^{-1}$, $R=10^{6}$ cm, and $B=10^{13}$ G the
maximum drop is a promisingly high $10^{21}$ V.  However, as in
models of pulsar radio emission, the true $\Phi_{\rm max}$ is
probably much less than $10^{21}$ V.  The basic problem is that
if the acceleration takes place along magnetic field lines with
finite radii of curvature, a seed electron accelerated by the
drop will emit curvature radiation.  If the curvature radiation
photons are energetic enough, then when they acquire a
sufficiently large angle to the magnetic field, they will
produce electron-positron pairs.  Each member of a pair is then
accelerated, but in opposite directions, and a pair cascade is
formed which rapidly shuts off the potential drop.  In both the
polar cap model (where the acceleration is at the stellar
surface; see, e.g., Ruderman \& Sutherland 1975) and the outer
gap model (where the acceleration takes place at a distance of a
few hundred stellar radii), the critical potential drop appears
to be a comparatively paltry $\sim 10^{12}-10^{13}$ V. The
critical drop rises with increasing radius of curvature, but
this requires a small area of emission.  It is thus likely that
the potential drop does not reach the required values.

Moreover, iron nuclei might not be accelerated to significantly
greater energies than protons are.  Binding energies in a strong
magnetic field are significantly greater than they are in zero
field (see, e.g., R\"osner et al. 1984; Miller \& Neuhauser
1991); in a $10^{12}$~G field at temperatures of $\sim 10^6$~K,
iron atoms are only partially ionized (about 3--4 times).  Their
energy after going through the potential drop would therefore be
only 3--4, and not 26, times that of protons, making
acceleration of iron nuclei to very high energies compared to
protons difficult, unless the electric field itself fully
ionized the atoms.

\subsection{Energy Loss Processes}

Energetic particles moving near a neutron star  experience various
energy losses that influence their propagation and emergent
energies.  When neutron stars are young, interactions of ions with
photons may be important, but for most of the life of a neutron star
we expect magnetic losses to dominate.

Young neutron stars are copious emitters of thermal X-ray
photons. If the surrounding photon density is large enough, then
cosmic rays accelerated near the surface will lose energy as
they move through this thermal bath of photons, through
processes such as inverse Compton scattering, photopion
production, and photodissociation.  These losses are significant
for surface temperatures in excess of $\sim 10^7$ K, implying
that they may hinder the production of high-energy cosmic rays
near neutron stars for a few years after the initial supernova.
Past this time, energy losses to these thermal photons are
minimal, and loss processes related to the stellar magnetic
field dominate.

Synchrotron radiation is the fundamental process to consider
when evaluating the effect of the magnetic field.  Electrons of
virtually any energy are constrained to follow field lines near
a neutron star, but because the synchrotron energy loss rate for
a given particle energy scales with the mass as $P_{\rm
synch}\sim m^{-4}$, protons or heavier ions may not follow field
lines, depending on the strength of the field. This also implies
that for UHECRs to emerge from the vicinity of a neutron star,
they should be accelerated several hundred radii from the star
to avoid synchrotron losses.  Moreover, synchrotron radiation
directly affects the relevance of other loss mechanisms such as
curvature radiation and trident pair production; curvature
radiation is significant only if synchrotron radiation forces
particles to follow magnetic field lines, and energy losses from
trident pair production are much less than those due to
synchrotron radiation, in our range of ion energies and magnetic
field strengths (see, e.g., Erber 1966).

Our final point in this section is that even if the magnetic
field is too weak to induce energy losses, e.g. for propagation
beyond $r_L$, it may still significantly affect the trajectories
of cosmic rays.  This will affect low-energy particles more than
high-energy particles, and, at a given energy, will delay the
emergence of heavier nuclei compared to protons.  This may lead
to a filter that selectively allows higher energy particles to
escape, and which affects the observed composition of particles
as a function of their energy.

\subsubsection{Synchrotron Radiation}

A particle of charge $q$ and mass $m$, with associated Lorentz factor
$\gamma$ and energy $E$, propagating at velocity $\beta c$ at an
angle $\alpha$ with respect to a magnetic field of strength $B$,
loses energy to synchrotron radiation at a characteristic rate
$${{\dot E}\over E}=-{2\over 3}r_0^2B^2\gamma\beta^2{\sin^2\alpha
\over mc}\; ,$$  where $r_0=q^2/mc^2$ is the classical radius of the
particle.  Thus, a nucleus of charge $Ze$ and mass $Am_p$ has a loss
rate of $${{\dot E}\over E}=3\times 10^5 {Z^4\over{A^3}}
\gamma\beta^2B_{12}^2 \sin^2\alpha\ {\rm s}^{-1}\; .$$ Note that for
a fixed energy $E$, $\gamma\sim 1/m$, and  ${\dot E}/E\sim m^{-4}$.

At a qualitative level, synchrotron losses are unimportant at a
radius $r$ from the neutron star when the characteristic energy
loss timescale at that radius exceeds the time to propagate a
distance $r$, if the cosmic ray is traveling in a straight line
away from the surface.  Assuming a dipolar magnetic field of
strength $B_0$ at the poles, if $\sin\alpha=1$, then protons or
iron nuclei with $\gamma=10^9$ suffer negligible synchrotron
losses for $B_{12}\simeq 10^{-6}$, which occurs at $r \simeq
10^8$ cm for a typical surface field $B_0 = 10^{12}$~G.  The
very strong dependence of the loss rate on distance ($\sim
r^{-6}$, for a dipole field) implies a transition radius $r \sim
10^8$ cm between regions of significant and insignificant
synchrotron losses, for straight line propagation.  In reality,
the trajectories of cosmic rays may be deflected by the magnetic
field, which increases the path length, but, given the steep
dependence of the loss rate on radius, we expect that for $r\gap
10^9$~cm the loss is insignificant.  Thus, $r\gap 10^9$~cm,
corresponding to acceleration at an outer gap or at the light
cylinder (for $P\gap 0.2$~s), is a promising location for
surviving UHECRs.

\subsubsection{Curvature Radiation}

Consider now the $r\lap 10^9$ cm region, where particles follow
field lines.  The most important energy loss process is then
curvature radiation (see, e.g., Sorrell 1987).  The power in
curvature radiation for a nucleus of charge $Z$ and Lorentz
factor $\gamma$ moving along field lines with radius of
curvature $R$ is $$P={2\over 3} {(Ze)^2c\over{R^2}}\gamma^4\;
.$$ If the power source is a constant electric field of
magnitude ${\cal E}=10^{12}{\cal E}_{12}$ V cm$^{-1}$, then the
energy at which the losses in curvature radiation equal the
power gain through the electric field is $$E_{\rm max} \approx
10^{18} \left(A\over{56}\right)\left(Z\over{26}\right)
^{-1/4}R_6^{1/2}{\cal E}_{12}^{1/4}\,{\rm eV}\; .$$ As expected,
the maximum energy rises with increased radius of curvature.
The radius of curvature at the stellar radius $R_*\approx 10^6$
cm of a dipole field line of maximum radius $R_{\rm max}\gg R_*$
is $$R\approx 2\left(R_{\rm max}R_*\right)^{1/2}\; .$$ Thus,
since the area of the polar cap is proportional to $R_{\rm
max}^{-1}$, the angular deviation of a field line from the
magnetic pole that just barely allows particles of energy
greater than $E_{\rm max}$ to escape scales as $\theta\sim
E_{\rm max}^{-2}$.  If particles are injected uniformly along
the polar cap, this implies a spectrum proportional to $E^{-5}$
with a sharp cutoff at the maximum potential drop.  If the
original source of energetic particles is at the neutron star
surface, this implies that either particle injection is strongly
biased toward the magnetic pole or there are further
acceleration mechanisms that harden the spectrum.

\subsubsection{Propagation outside the light cylinder}

Outside the light cylinder, conservation of magnetic flux
implies that the magnetic field scales as $B\sim r^{-1}$. Since
the radial component of the magnetic field scales as $B_r\sim
r^{-2}\, $ far from the light cylinder, the azimuthal component
dominates.  Therefore, most nuclei traveling away from the star must
eventually cross field lines.  If the gyration radius of a
particle $r_g$ at the light cylinder is less than $r_L$, the particle
might become trapped. This may lead to collisionless shocks or
other mechanisms by which energy can be redistributed amongst
the particles.  Note that since
$${r_g\over{r_L}}=0.03 \;
E_{15}B_{0,12}^{-1}\left(Z\over{26}\right)^{-1} P_{-1}^2\, ,$$
this can also act as a filter which preferentially allows higher
energy particles to escape. Note also that for a given energy of
the particle, low-$Z$ species escape sooner than high-$Z$
species. For example, iron nuclei of energy $10^{15}$~eV are
trapped only up to a neutron star period of $P\sim 0.6$ s,
whereas protons escape for periods greater than about 0.1
s. Alternatively, for a given neutron star period or stage in
the star's lifetime, protons need less energy to overcome this
effect than do iron nuclei.

\section{Conclusions}

In this paper, we have examined the case for generating UHECRs
from isolated neutron stars.  From overall power and energy loss
criteria, we have strongly constrained possible models.  We find
that if isolated neutron stars produce most UHECRs, the
generation mechanism must be ultimately powered by rotation and
the particles must come from near the star rather than from the
interstellar medium.  We also find that potential drops along
magnetic fields cannot accelerate particles to energies above
$10^{15}$ eV, since electron-positron pair cascades are created
that significantly reduce the available energies.  Unless the
acceleration takes place farther than $\sim 10^9$~cm from the
star, synchrotron losses are likely to dissipate a significant
fraction of the particle energy. 

High-energy gamma rays from pulsars also provide evidence
against a substantial fraction of the rotational energy of
neutron stars being converted to particles of Lorentz factor
$\gap$ 10$^7$ within the light cylinder. These objects have a
gamma-ray luminosity that is only a small fraction of their
total spin-down power, typically less than 10\% (Nel et
al. 1996, and references therein). This is contrary to what
would be observed if energetic particles were accelerated close
to rapidly rotating, strongly magnetized neutron stars (as we
discuss in \S 4).  Thus, if UHECRs come from isolated neutron
stars, the acceleration region must be near or outside $r_L$.
Finally, we note that if acceleration does indeed take place
beyond the light cylinder, the azimuthal nature and radial
dependence of the magnetic field in this region may form a
magnetic bottle that preferentially lets out higher energy
particles.

We thank M. Ruderman for pointing out the relevance of
multiwavelength observations of pulsars. We also thank G. Sigl,
R. Epstein, and K. Green for useful references and discussions.
A. V. and A. O. were supported in part by the DOE through grant
DE-FG0291 ER40606. M. C. M. gratefully acknowledges the support
of a {\it Compton GRO} Fellowship, and NASA grants NAG 5-2868 and
5-2687.

\def\nature{{\rm Nature}} \def\nucphys{{\rm Nuc. Phys.}}
\def\nucphysa{{\rm Nuc. Phys. A}} \def\physletb{{\rm Phys. Lett. B}}
\def\physrevc{{\rm Phys. Rev. C}} \def\physrevd{{\rm Phys. Rev. D}}
\def\sovphysjetp{{\rm Soviet~Phys.~JETP}} \def\ptpl{{\rm
Progr.Theor.Phys.Lett}} \def\ptps{{\rm Prog.Theor.Phys.Suppl.}}
\def\ptp{{\rm Prog. Theor. Phys.}}

\begin{center}
{\large \bf{REFERENCES}}  \\
\end{center}

\noindent Axford, W. I. 1994, Ap.J.Supp., 90, 937. \\
Becker, R. H. \& Helfand, D. J. 1983, Nature, 302, 688. \\
Biermann, P. L. 1993, A \& A, 271, 649.\\
Bildsten, L., Salpeter, E. E. \& Wasserman, I. 1992, Ap.J. 384,
143. \\
Bird, D. J. et al. 1994, Ap.J., 424, 491. \\
Blaes, O. \& Madau, P. 1993, Ap.J., 403, 690 (BM93). \\
Blaes, O., Warren, O. \& Madau, P. 1995, Ap.J., 454, 370.\\
Bondi, H. \& Hoyle, F. 1944, M.N.R.A.S., 104, 273. \\
Cheng, K. S., Ho, C., Ruderman, M. A. 1986, Ap.J., 300, 500.\\
Erber, T. 1966, Rev. Mod. Phys., 38, 626. \\
Frail, D. A., Goss, W. M. \& Whiteoak, J. B. Z. 1994, Ap.J., 437, 781.\\
Gaisser, T. K. et al. 1993, Phys. Rev. D, 47, 1919. \\
Harding, A. K. 1990, Nucl. Phys. B (Proc. Supp.), 14A, 3. \\
Illarionov, A., \& Sunyaev, R. 1975, A \& A, 39, 185. \\
Lyne, A. G. \& Lorimer, D. R. 1994, Nature, 369, 127. \\
Madau, P. \& Blaes, O. 1994, Ap.J., 423, 748 (MB94). \\
Milgrom, M. \& Usov, V. 1996, Astropart. Phys., 4, 365. \\
Miller, M. C. \& Neuhauser, D. 1991, M.N.R.A.S., 253, 107. \\
M\"uller, D. et al. 1991, Ap.J., 374, 356. \\
Nel, H. I. et al. 1996, Ap.J., 465, 898. \\
R\"osner, W., Wunner, G., Herold, H. \&  Ruder, H. 1984,
J. Phys. B: At. Mol. Phys., 17, 29. \\
Ruderman, M. A. \& Sutherland, P. G. 1975, Ap.J., 196, 51. \\
Shapiro, S. L. \& Teukolsky, S. A. 1983, Black Holes, White Dwarfs, and
Neutron Stars (New York: Wiley), 278. \\
Shemi, A. 1995, M.N.R.A.S., 275, 115. \\
Sorrell, W. H. 1987, Ap.J., 323, 647. \\
Timmes, F. X., Woosley, S. E. \& Weaver, T. A. 1996, Ap.J., 457,
834. \\
Vietri, M. 1996, M.N.R.A.S., 278, L1.

\end{document}